\documentclass[aps,pre, preprint, groupedaddress]{revtex4-1}

\usepackage{times}

\usepackage{stmaryrd}

\usepackage{amsmath, amsthm, amsfonts}
\usepackage{graphicx}    
\usepackage{color}          
\usepackage{subfigure}   
\usepackage{hyperref}     
\usepackage{graphics,xcolor,amssymb,cases,setspace,
setspace,lipsum,dsfont,textcomp}
\usepackage[english]{babel}

\begin{document}

\title{Visco--thermal effects in acoustic metamaterials: from total transmission to total reflection and high absorption}
\author{Miguel Moler\'on, Marc Serra-Garcia, Chiara Daraio}

\affiliation{Department of Mechanical and Process Engineering, Swiss Federal Institute of Technology (ETH), Zurich, Switzerland}

\date{\today}

\begin{abstract}

We theoretically and experimentally investigate visco--thermal effects on the acoustic propagation through metamaterials consisting of rigid slabs with subwavelength slits embedded in air. We demonstrate that this unavoidable loss mechanism is not merely a refinement, but it plays a dominant role in the actual acoustic response of the structure. Specifically, in the case of very narrow slits, the visco--thermal losses avoid completely the excitation of Fabry--Perot resonances, leading to 100\% reflection. This is exactly opposite to the perfect transmission predicted in the idealised lossless case. Moreover, for a wide range of geometrical parameters, there exists an optimum slit width at which the energy dissipated in the structure can be as high as 50\%. This work provides a clear evidence that visco--thermal effects are necessary to describe realistically the acoustic response of locally resonant metamaterials. 

\end{abstract}
\pacs{}

\maketitle

\section{Introduction} Metamaterials are artificial structured materials in which the presence of resonances in the micro/meso-scale leads to unprecedented properties \cite{Craster2012, Deymier2013}. In recent years, metamaterials consisting of rigid slabs with subwavelength perforations have attracted considerable attention due to their ability to achieve normalised-to-area transmission (\textit{i.e.}, transmission normalised to the fraction of area occupied by the holes) significantly bigger than unity, a phenomenon known as extraordinary acoustic transmission (EAT) \cite{Lu2007}. This phenomenon, analogue to extraordinary optical transmission \cite{Ebbesen1998}, can be achieved by means of different physical mechanisms, such as the excitation of Fabry--Perot (FP) resonances in the holes \cite{Lu2007,Christensen2008,Wang2010,Zhou2010}, the acoustic Brewster angle \cite{DAguanno2012, Maurel2013}, or the acoustic analog to the supercoupling effect in density--near--zero metamaterials \cite{Fleury2013}. Promising applications to this fascinating phenomenon have been suggested, including acoustic collimators \cite{Christensen2007}, superlenses \cite{Zhu2011}, highly efficient Fresnel lenses \cite{Moleron2014}, beam shifters \cite{Wei2015}, passive phased arrays \cite{Li2015} and invisibility cloaks \cite{Zhao2015}. 

A main limitation in the practical realization of EAT and other unconventional phenomena in locally resonant metamaterials arises from the unavoidable presence of viscous and thermal boundary layers at the solid-fluid interface \cite{Kirschhoff1868, Rayleigh1901}, which can induce important losses. However, only a few papers have investigated boundary layer effects in metamaterials. In Ref. \cite{Theocharis2014}, it was demonstrated that visco--thermal dissipation has a strong influence in the slow sound propagation in waveguides with side resonators, hindering the formation of near--zero group velocity dispersion bands. This feature was exploited later to design low frequency acoustic absorbers \cite{Groby2015}. More recently, visco--thermal dissipation in microslits has been used to enhance the attenuation of metamaterials \cite{Ruiz2015}, and important boundary layer effects have also been reported in phononic crystals \cite{Duclos2009, Guil2014}.

The goal of the present work is to investigate visco--thermal losses in acoustic metamatarials consisting of rigid slabs with subwavelength slits. Previous studies have already proven that this dissipation mechanism may significantly attenuate the otherwise perfect transmission peaks associated to FP resonances \cite{Ward2015, DAguanno2012}, while the nonresonant EAT mechanism based on the Brewster angle remains much less affected \cite{ DAguanno2012}. However, these works lack a clear theoretical analysis of the behaviour of the system in the presence of losses and do not describe completely the physical mechanisms governing the reduction of transmission, in relation to reflection and/or dissipation. This paper complements these earlier studies, providing an experimental and theoretical analysis of the acoustic transmission, reflection, and absorption in the presence of visco--thermal dissipation. Our results demonstrate that this loss mechanism avoids completely the excitation of Fabry--Perot resonances in gratings with very narrow slits, which leads to 100\% reflection. In addition, we prove that there is an optimum slit width that maximises the acoustic absorption, which reaches more than 50\%.

\section{Experimental Setup} 
\begin{figure}[t!]
\centering
\includegraphics[width=12cm]{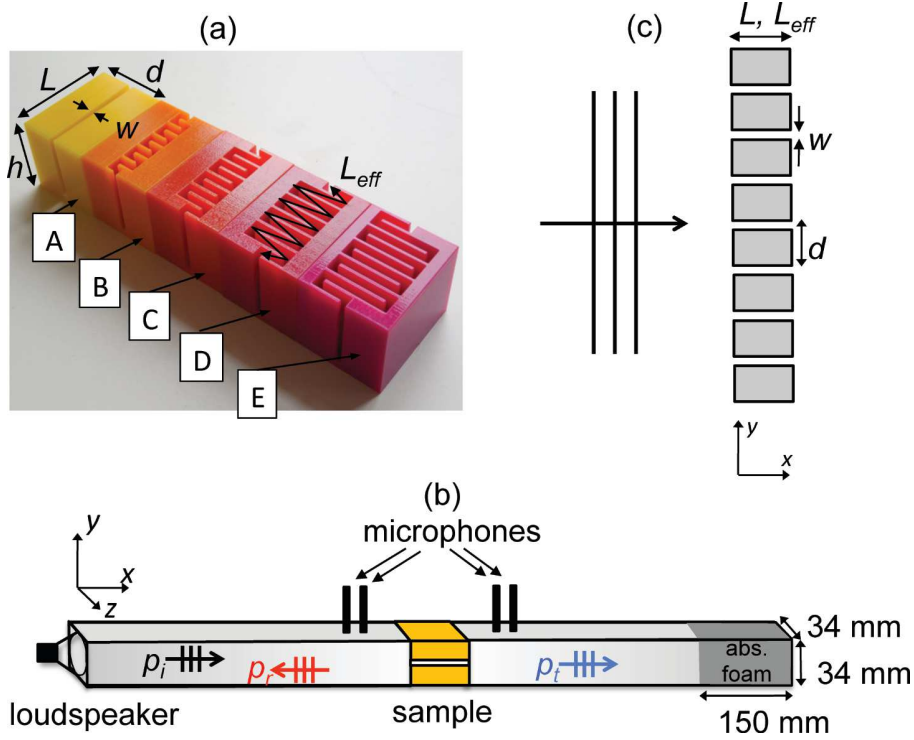}
\caption{(a) Tested under test, characterised by the slit width $w=2.7$~mm, the grating period $d=34$~mm and the effective grating thickness $L_{eff}=L=52$~mm for Sample A, $L_{eff}=2.08L$ for Sample B, $L_{eff}=3.16L$ for Sample C, $L_{eff}=4.24L$ for Sample D and $L_{eff}=5.32L$. The cover of the samples has been removed to reveal the internal structure. (b) Schematic of the experimental setup. (c) 2D perforated slab equivalent to the one studied experimentally.}
\label{Fig1}
\end{figure}

Figure \ref{Fig1}(a) shows the samples under experimental consideration, which were fabricated using 3D printing (Stratasys Objet500). The material used was a rigid thermoplastic (Vero materials (c)) with manufacturer specified mass density 1.17--1.18 g/cm$^3$ and modulus of elasticity 2--3 GPa. The samples are rectangular blocks with an air channel connecting the input and output sides. Sample A has a straight channel, while in samples B to E the channels describe a zigzag path. For wavelengths much bigger than the height of the corrugations, the zigzag channel behaves similarly to a straight slit with effective length $L_{eff}$, which is approximately equal to shortest path taken by the wave to pass through the structure \cite{Liang2012, Li2013}. The samples were placed between two aluminium tubes with square cross-section and 34~mm inner side, as shown in Figure \ref{Fig1}(b). The square cross-section artificially imposes periodic boundary conditions in the transverse directions. Since, from the point of view of the plane waves traveling inside the tubes, the samples' geometry is constant along the $z-$direction (see Fig.~\ref{Fig1}(b)), the structure is equivalent to a 2D rigid slab with a periodic array of slits along the $y-$direction, as illustrated in Fig.~\ref{Fig1}(c). The relevant geometrical parameters of this equivalent system are the slit width $w=2.7$~mm, the grating period $d=34$~mm and the effective grating thickness $L_{eff}$, $L_{eff}=L=52$~mm for Sample A, $L_{eff}=2.08L$ for Sample B, $L_{eff}=3.16L$ for Sample C, $L_{eff}=4.24L$ for Sample D and $L_{eff}=5.32L$ for Sample E. The transmission $T=|p_t/p_i|^2$, reflection $R=|p_r/p_i|^2$, and absorption coefficient $A=1-R-T$ were measured with 4 microphones (G.R.A.S. 40BD) using the two-port technique \cite{Abom1991}, where  $p_i, p_r$ and $p_t$ are respectively the complex amplitude of the incident, reflected and transmitted plane mode [see Fig.~\ref{Fig1}(b)]. We measured these quantities using phase sensitive detection with a sinusoidal wave as reference signal, injected to a loudspeaker (Clarion SRE 212H) on the left extremity. A 150~mm thick absorbing foam was placed on the right side to minimise backward reflections. The testing frequency range was limited to $[1-5]$~kHz. The lower limit is imposed by the loudspeaker, which is not able to radiate sound below approximately 1~kHz. The upper limit is imposed by the cutoff frequency of the first high-order mode in the ducts, approximately 5~kHz, so that only the plane mode excites the samples. 

\section{Model} The acoustic propagation through the grating depicted in Fig.~\ref{Fig1}(c) is modelled using a multimodal approach developed in previous works by the authors \cite{Moleron2014, Moleron2015}. We express the acoustic pressure field, $p(x,y)$, as a modal decomposition,
\begin{equation}\label{GenSol}
p(x,y) =  \sum_{n} \left( A_n  \text{e}^{\jmath \beta_n x}  +  B_n  \text{e}^{-\jmath \beta_n x}  \right) \phi_n(y),
\end{equation}
where $A_n$ and $B_n$ are respectively the modal amplitude of the n$-th$ forward and backward mode, $\beta_n$ are the propagation constants, and $\phi_n(y)$ are the eigenfunctions. In the surrounding space with periodic boundary conditions, the eigenfunctions are
\begin{equation}\label{phifree}
\phi_n(y) = \frac{1}{\sqrt{d}}\text{e}^{\jmath[2n\pi/d+k\sin(\theta)] y}, n\in\mathds{Z},
\end{equation}
where $k=\omega/c$ is the wavenumber in free space, $\omega$ is the angular frequency, $c$ is the sound speed, and $\theta$ is the incidence angle with respect to $x$ of the impinging plane wave (here $\theta=0$). Assuming rigid boundaries, the eigenmodes of the slit are given by,
\begin{equation}\label{phirigid}
\phi_n(y) = \sqrt{\frac{2-\delta_{n,0}}{w}} \cos\left[\frac{n\pi}{w} \left(y-\frac{w}{2}\right) \right], n\in\mathds{N}
\end{equation}
with $\delta_{n,0}$ the Kronecker delta ($\delta_{n,0}=1$ for $n=0$ and $\delta_{n,0}=0$ otherwise).

In the absence of losses, the propagation constants of the slit modes are given by the dispersion relation $\beta_n^2=k^2-(n\pi/w)^2$. The effect of the viscous and thermal losses can be taken into account by introducing an additional term into these propagation constants (see Ref. \onlinecite{Bruneau1987}),
\begin{equation}\label{Eqbn}
\beta_n^2 = k^2 - \left( \frac{n\pi}{w}\right)^2 +\frac{2k}{w}(2-\delta_{n,0})\left(\text{Im}\{\varepsilon_n\} -\jmath\text{Re}\{\varepsilon_n\} \right)
\end{equation}
where

\begin{equation}\label{en}
\varepsilon_n=\left[1 - \left( \frac{n\pi}{wk} \right)^2 \right]\varepsilon_v + \varepsilon_t,
\end{equation}

\begin{equation}\label{ev}
\varepsilon_v = (1+\jmath)\sqrt{\frac{k l_v}{2}},
\end{equation}
and
\begin{equation}\label{et}
\varepsilon_t = \frac{(1+\jmath)}{(\gamma - 1)}\sqrt{\frac{k l_t}{2}},
\end{equation}
In Eqs.\eqref{en}--\eqref{et}, $\gamma=1.4$ the adiabatic specific heat ratio of air, $l_v$ is the viscous characteristic length and $l_t$ is the thermal characteristic length. At standard conditions, $c\approx 344$~m/s, $l_v$ and $l_t$ are respectivelly $l_v=4.5\times10^{-8}$~m and $l_t=6.2\times10^{-8}$~m (see Ref. \onlinecite{Bruneau1987}). We note that, according to the time convention chosen in this paper ($\text{e}^{-\jmath\omega t}$), we only keep solutions to \eqref{Eqbn} fulfilling $\text{Re}\{\beta_n\}, \text{Im}\{\beta_n\}\geqslant0$.

Writing the continuity equations of pressure and normal velocity at the interface between the grating and the surrounding space leads to the reflection and transmission matrices, $\mathbf{R}$ and $\mathbf{T}$ (see \cite{Moleron2014, Moleron2015} for details), defined as $\vec{A}_R=\mathbf{R}\vec{A}_I$ and $\vec{A}_R=\mathbf{T}\vec{A}_I$, where $\vec{A}_I$, $\vec{A}_R$ and $\vec{A}_T$ are row vectors containing the incident, reflected and transmitted modal amplitudes.  Finally, the energy reflection, transmission and absorption coefficients are given respectively by

\begin{equation}\label{Rnum}
R=\frac{\text{Re}\left\{ {\vec{A}^\text{t}_R}(\mathbf{Y}{\vec{A}_R} )^* \right\}} {\text{Re}\left\{ {\vec{A}^\text{t}_I}(\mathbf{Y}{\vec{A}_I} )^* \right\} },
\end{equation}

\begin{equation}\label{Tnum}
T=\frac{\text{Re}\left\{ {\vec{A}^\text{t}_T}(\mathbf{Y}{\vec{A}_T} )^* \right\} } {\text{Re}\left\{ {\vec{A}^\text{t}_I}(\mathbf{Y}{\vec{A}_I} )^* \right\} },
\end{equation}

\begin{equation}
A=1-R-A,
\end{equation}
where $\mathbf{Y}=\text{diag}\{ \beta_n/\rho c k \}$ and superscripts "t" and "$\ast$" indicate respectively the transpose and the complex conjugate. The series of Eq.~\eqref{GenSol} was truncated to 40 modes in the free, periodic space and 5 modes in the slit, from which only the fundamental one [$n=0$ in Eqs. \eqref{phifree}--\eqref{Eqbn}] is propagative.

\section{Results} 
\begin{figure}[h!]
\centering
\includegraphics[width=12cm]{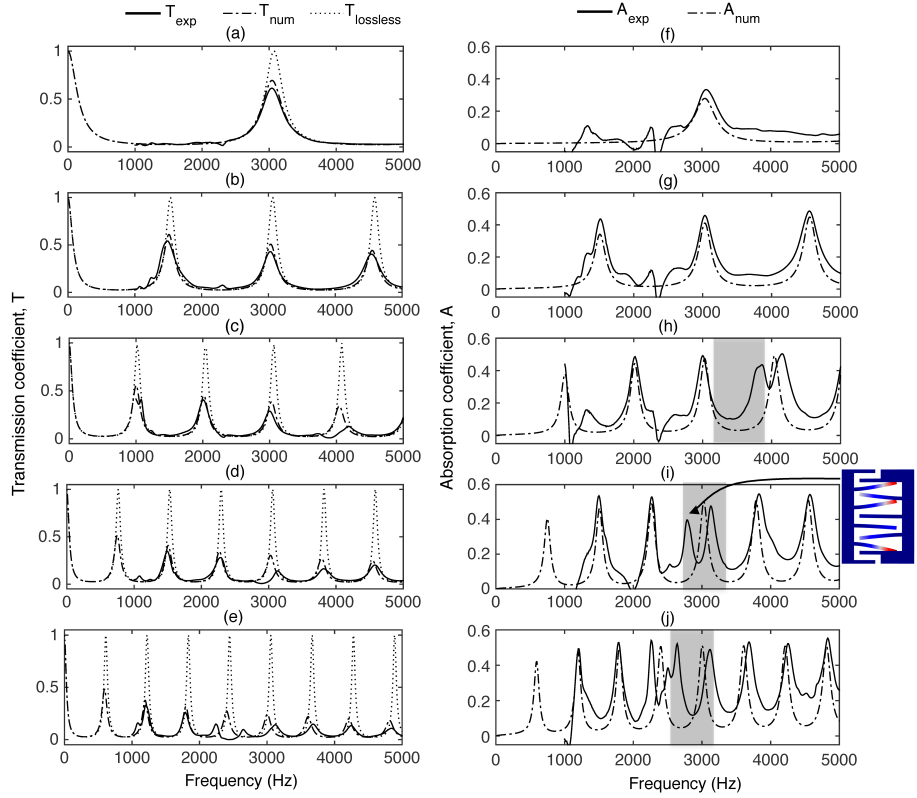}
\caption{(a) to (e) show the transmission coefficients for samples A to E, respectively. (f) to (j) show the absorption coefficients for samples A to E, respectively. The shaded regions in (h), (i) and (j) represent the frequency range in which vibrational modes may exist. An example of these modes at $f=2750$~Hz is shown in (i).}
\label{Fig2}
\end{figure}

We start our analysis by studying the influence of $L_{eff}$ in the acoustic response of the samples. Figures \ref{Fig2}(a)--(e) show the experimental (solid lines) and numerical (dash-doted lines) transmission coefficients. The lossless transmission coefficients (doted lines), obtained by replacing $\beta_n$ with $\beta_n^2=k^2-(n\pi/w)^2$ in Eq.~\eqref{Eqbn}, are also shown. The lossless transmission coefficients exhibits the perfect transmission peaks typical of FP resonances at $f\approx sc/2L_{eff}$, with $s$ a positive integer. However, when visco--thermal effects are included in the model, we observe a strong attenuation of the resonance peaks as $L_{eff}$ increases, which is in good agreement with the experimental results. We also observe a downshift of the resonance frequencies compared to the lossless case, due to the slowing down of the wave because of the dissipation \cite{Ward2015}. These figures represent a first and clear evidence that neglecting visco--thermal effects leads to a poor description of the actual metamaterial response.


Figures \ref{Fig2}(f)-(j) show the absorption coefficients as a function of frequency for the different samples. Experimental and numerical results are in good qualitative agreement and demonstrate a strong dissipation at the FP resonance frequencies. The attenuation peaks reach between 33\% and 55\% in experiments, and between 30\% and 50\% in the numerical results. We also notice an increase of absorption with frequency, which is due to the fact that $\text{Im}\{\beta_0\}$ ($\beta_0$ is the propagation constant of the fundamental slit mode) also increases with frequency, approximately as $\sqrt{f}$, see Eq.~\eqref{Eqbn}. 

Although the agreement between experimental and numerical results is globally good, particularly in terms of the amplitude of the absorption peaks, we also observe some discrepancies. The experimental absorption coefficient is in general higher than the numerical one, which can be atributed to additional losses in the experimental setup, as visco--thermal losses in the aluminium tubes (these effects are only modelled within the slits), material losses, or losses due to energy leakage between the different pieces forming the experimental apparatus. We also observe features in the experimental curves that are not observed in the numerical results. The experimental curves around the fourth peak in Figs.~\ref{Fig2}(h)-(j) exhibit an additional peak, not observed in the numerical curve. The origin of these peaks is the excitation of vibrational modes of the samples, which is consistent with the downshift of the peak frequencies as the height of the internal corrugations increases. We verified this through the computation of the samples' vibrational modes using Comsol Multiphysics. The shaded regions in Figs.~\ref{Fig2}(h)--(j) represent the frequency range in which vibrational modes were obtained, accordingly to the range of mechanical parameters provided by the manufacturer. An example of these modes at $f=2750$~Hz is displayed in Fig.~\ref{Fig2}(i).

It is remarkable to achieve such high absorption using visco--thermal effects, considering that the slit width is about 2 orders of magnitude bigger than the viscous and thermal boundary layers ($\sim10^{-5}$m \cite{Morse}). This behaviour, which is consistent with recent observations by Ward \textit{et al.} \cite{Ward2015}, is rather surprising as intuition suggest that visco--thermal effects would only become relevant when the slit width is of the same order as the boundary layers' thickness.

\begin{figure}
\centering
\includegraphics[width=12cm]{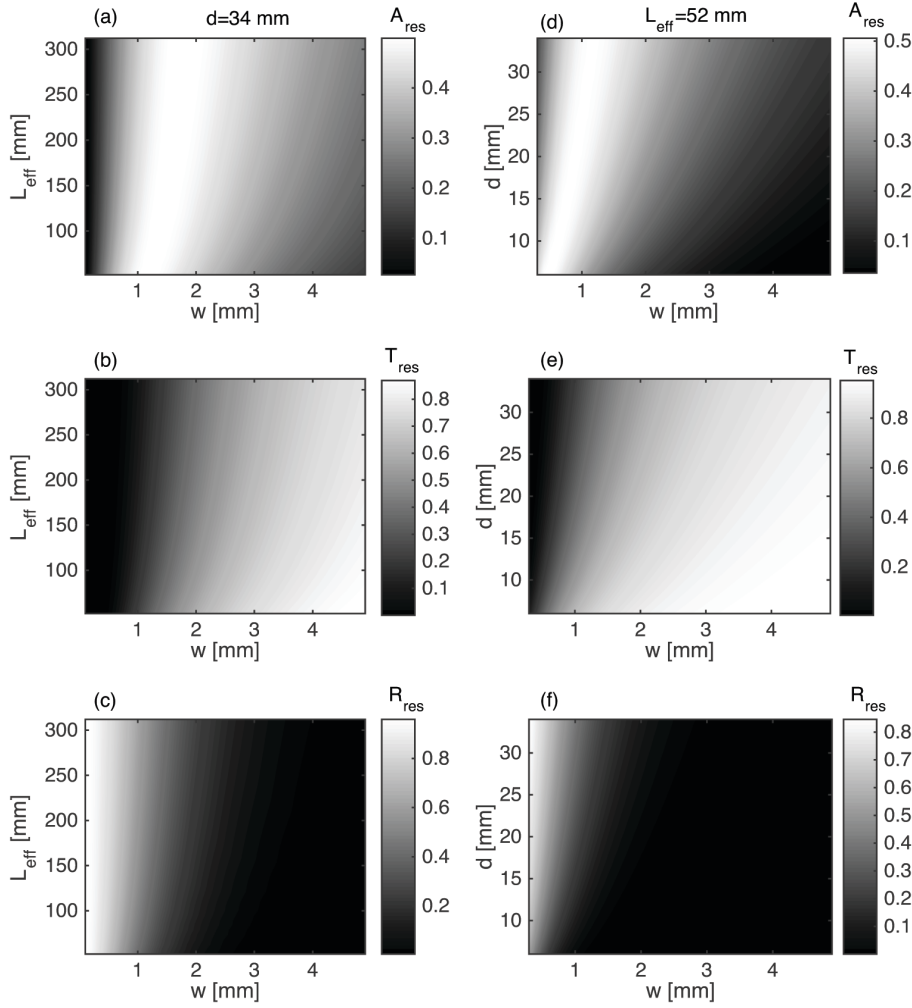}
\caption{(a) numerical absorption, (b) transmission and (c) reflection coefficients at resonance as a function of $L_{eff}$ and $w$, for a constant $d=34$~mm. (d) Numerical absorption, (e) transmission and (f) reflection coefficients at resonance as a function of $d$ and $w$, for a constant $L_{eff}=52$~mm.}
\label{Fig3}
\end{figure}

To quantify the maximum amount of energy that can be absorbed in these structures by the combination of FP resonances and visco--thermal losses, we have computed the amplitude of the first absorption peak, $A_{res}$, as a function of the geometrical parameters. Fig.~\ref{Fig3}(a) shows $A_{res}$ versus $L_{eff}$ and $w$ for $d=34$~mm fixed, and Fig.~\ref{Fig3}(d) shows the same quantity versus $d$ and $w$ for $L_{eff}=52$~mm fixed. We observe that the parameter having a stronger influence on the response is the slit width $w$. For any $L_{eff}$ and $d$ considered, $A_{eff}$ exhibits a maximum between $w=0.5$~mm and $w=2$~mm, at which the structure dissipates about 50\% of the incident energy. 

The vanishing of $A_{res}$ after its maximum is due to the fact that the acoustic impedance of the surrounding media, $Z_1=\rho c / d$,  approaches that of the slit cavity, $Z_2=\rho c k/w\beta_0$,  as $w\to d$. This inhibits the formation of a high amplitude standing wave in the slit cavity and reduces the ability to dissipate energy. However, the vanishing of $A_{res}$ as $w\to 0$ is less straightforward. In principle,  visco--thermal losses (given by $\text{Im}\{\beta_0\}$) and the strength of the FP resonance (provided by the impedance ratio $Z_1/Z_2$) increase as $w\to 0$ when considered separately. Hence, one should expect the dissipation to increase also in this region. However, inspecting the transmission and reflection coefficients at resonance, respectivelly $T_{res}$ [Figs.~\ref{Fig3}(b) and \ref{Fig3}(e)] and $R_{res}$  [Figs.~\ref{Fig3}(c) and \ref{Fig3}(f)], we observe that the structure behaves as a perfect reflector as $w\to 0$, meaning that very little energy is stored (and therefore dissipated) in the resonator. Remarkably, although the dramatic drop in the transmission is a direct consequence of the presence of losses in the system, this drop is not reflected in an increase of dissipation. 

To explain this counterintuitive behaviour, we derive analytical expressions for the reflection and transmission coefficients. To accomplish this, we reduce our model to only the fundamental mode in both the slit cavity and the free, periodic space [$n=0$ in Eq.~\eqref{GenSol}]. The amplitude reflection and transmission coefficients of this mode, respectively $r_0$ and $t_0$, take the following form
\begin{equation}\label{r0}
r_0= \frac{ \left( 1- \frac{Z_1^2}{Z_2^2}  \right)\left( \text{e}^{-2\jmath \beta_0 L_{eff}} - 1\right)}{\left( \frac{Z_1}{Z_2} +1\right)^2 \text{e}^{-2\jmath \beta_0 L_{eff}} -\left( \frac{Z_1}{Z_2} -1\right)^2}
\end{equation}
and 
\begin{equation}\label{t0}
t_0=\frac{4\frac{Z_1}{Z_2}\text{e}^{\jmath \beta_0 L_{eff}} }{\left(1+\frac{Z_1}{Z_2}\right)^2 -\left( \frac{Z_1}{Z_2} -1 \right)^2\text{e}^{2 \jmath \beta_0 L_{eff}} },
\end{equation}
from which $R$ and $T$ are obtained as $R=|r_0|^2$ and $T=|t_0|^2$. In the absence of losses ($\beta_0=k$) the FP resonances appear when $\text{e}^{2\jmath \beta_0 L_{eff}}=1$, or equivalently when $k=s\pi/L_{eff}$, which leads to $R=0$ and $T=1$, regardless of $L_{eff}$, $w$, and $d$  \cite{Lu2007,Christensen2008,Wang2010,Zhou2010, Li2013}. However, when visco--thermal losses are accounted for, $\beta_0$ is $\beta_0=\text{Re}\{\beta_0\}+\jmath\text{Im}\{\beta_0\}$, and FP resonances appear when $\text{e}^{2\jmath \text{Re}\{\beta_0\} L_{eff}}=1$. In such case, the reflection and transmission coefficients become
\begin{equation}\label{Rloss}
r_{0,res}= \frac{ \left( 1- \frac{Z_1^2}{Z_2^2}  \right)\left( \text{e}^{2 \text{Im}\{\beta_0\} L_{eff}} - 1\right)}{\left( \frac{Z_1}{Z_2} +1\right)^2 \text{e}^{2 \text{Im}\{\beta_0\} L_{eff}} -\left( \frac{Z_1}{Z_2} -1\right)^2}
\end{equation}
and
\begin{equation}\label{Tloss}
t_{0,res}=\frac{4\frac{Z_1}{Z_2}\text{e}^{- \text{Im}\{\beta_0\} L_{eff}} }{\left(1+\frac{Z_1}{Z_2}\right)^2 -\left( \frac{Z_1}{Z_2} -1 \right)^2\text{e}^{-2 \text{Im}\{\beta_0\} L_{eff}} }.
\end{equation}
Contrary to the conservative case [Eqs. \eqref{r0} and \eqref{t0}], we see that the acoustic response at resonance depends on the geometrical parameters, both on $L_{eff}$ and $w/d$ (through the impedance ratio $Z_1/Z_2\propto w/d$). For $w\ll d$, that is $Z_1/Z_2 \to 0$, one has $R\to 1$ and $T\to 0$, which is consistent with the numerical results in Fig.~\ref{Fig3}. Remarkably,  this result holds for any $L_{eff}>0$ as long as dissipation is present in the slit ($\text{Im}\{\beta_0 \}>0$), which is always true in realistic situations. In other words, the reflection always tends to 1 in slabs such that $w\ll d$, regardless of the thickness $L_{eff}$. This is visible in Fig.~\ref{Fig3}(c). Another implication of Eqs. \eqref{Rloss} and \eqref{Tloss} is that, when $w$ is very small (say $w<1$~mm), high transmission can be achieved only if the period $d$ is comparable to $w$, that is when $Z_1/Z_2 \to 1$. This means that EAT (\textit{i.e.} transmission considerably bigger than unity when normalised to the ratio $w/d$) cannot be achieved in slabs with very narrow slits. 
 
\begin{figure}
\centering
\includegraphics[width=12cm]{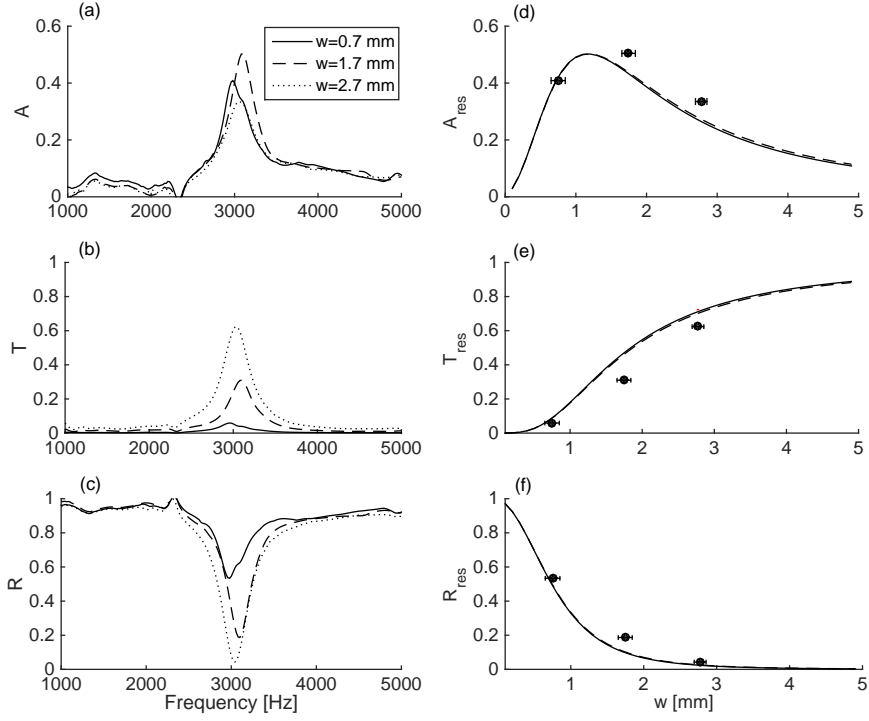}
\caption{(a) Experimental absorption, (b) transmission and (c) reflection coefficient for sample A with slit width 0.7~mm (solid line), 1.7~mm (dashed line) and 2.7~mm (dotted line). (d) Absorption, (e) transmission and (f) reflection coefficient at resonance as a function of $w$. Dots represent the experimental data, obtained from the maxima in figure (a) and (b) and the minima in figure (c). Solid lines represent analytical results [Eqs.~\eqref{Rloss} and \eqref{Tloss}] and dashed lines represent numerical results [Eqs.~\eqref{Rnum} and \eqref{Tnum}]. }
\label{Fig4}
\end{figure}

In order to obtain an experimental evidence for this behaviour we have fabricated and tested two additional samples. The new samples are identical to sample A, but the slit width is equal to 0.7~mm and 1.7~mm. Figures~\ref{Fig4}(a) to \ref{Fig4}(c) show, respectively, the experimental absorption, transmission and reflection coefficients. These figures confirm the behaviour described previously in Fig.~\ref{Fig3}: the reflection (transmission) increases (decreases) monotonically as $w\to 0$, and there is an optimum $w$ that maximises the absorption. For a quantitative comparison of experiments with theory, Figs.~\ref{Fig4}(d)--\ref{Fig4}(f) show, respectively, $A_{res}$, $R_{res}$ and $T_{res}$ as a function of $w$. Solid lines represent the analytical results obtained with Eqs.~\eqref{r0} and \eqref{t0}. Dashed lines represent the numerical result obtained with the multimodal method, Eqs.~\eqref{Rnum} and \eqref{Tnum}. The experimental data, obtained from the maxima of $A$ and $T$, or minima of $R$ in Figs.~\ref{Fig4}(a)-(c) is represented with dots.     
Horizontal error bars in experimental results represent the standard deviation of the actual slit width from the desired values, measured at four different points along the slit. This deviation was less than 0.1~mm for all samples. The trend exhibited by experimental results agrees very well with both numerical and analytical results, either in absorption, transmission and reflection, which corroborates the theoretical predictions.

\section{Conclusion} In summary, visco--thermal effects are essential to describe realistically the acoustic response of metamaterials composed of rigid slabs with subwavelength slits. Due to the presence of this loss mechanism, the behaviour of the structure at the FP resonances depends completely on the geometrical parameters, which can be adjusted to achieve high transmission, high absorption or high reflection. Our work may have important implications in the design of acoustic metamaterials. For instance, the inability to obtain sharp FP resonances in slabs with very narrow slits compromises the practical realization of resonant EAT at ultrasonic frequencies. On the other hand, understanding and exploiting this property gives the possibility to design subwavelengh sized, tailorable devices providing high transmission, high reflection or high absorption. From a more general perspective, we expect that our work will result in widespread consideration of this unavoidable loss mechanism in acoustic metamaterials research.


%

\end{document}